# Invariance of the spark, NSP order and RIP order under elementary transformations of matrices

Jiawang Yi[1,2], Guanzheng Tan[1]

[1]School of Information Science and Engineering, Central South University, Changsha 410083, China.

[2]School of Computer & Communication Engineering, Changsha University of Science & Technology, Changsha 410114, China.

Emails: {JiawangYi, tgz}@csu.edu.cn

**Abstract:** The theory of compressed sensing tells us that recovering all $k$-sparse signals requires a sensing matrix to satisfy that its spark is greater than $2k$, or its order of the null space property (NSP) or the restricted isometry property (RIP) is $2k$ or above. If we perform elementary row or column operations on the sensing matrix, what are the changes of its spark, NSP order and RIP order? In this paper, we study this problem and discover that these three quantitative indexes of sensing matrices all possess invariance under all matrix elementary transformations except column-addition ones. Putting this result in form of matrix products, we get the types of matrices which multiply a sensing matrix and make the products still have the same properties of sparse recovery as the sensing matrix. According to these types of matrices, we made an interesting discovery that sensing matrices with deterministic constructions do not possess the property *universality* which belongs to sensing matrices with random constructions.

**Keywords**：Compressed sensing, Spark, Null space property, Restricted isometry property, Invariance, Elementary transformation.

## 1. Introduction

The idea of compressed sensing or compressive sensing (CS) [5, 7, 12] is to attain the goal of dimensionality reduction by exploiting the sparsity or compressibility of signals. A signal $x$ is defined as $k$-sparse if it has at most $k$ nonzero entries, and the set of all $k$-sparse signals is denoted by $\Sigma_k$. Those signals which can be well-approximated by a sparse signal are called compressible or approximately sparse signals. The realization of CS or sparse recovery requires the designing of a measurement system

$$y = Ax$$

which can recover a sparse or approximately sparse signal $x \in R^n$ from a vector $y \in R^m$ made up of $m$ measurements, where $m$ is typically much smaller than $n$. The $m \times n$ matrix $A$ is referred to as a sensing matrix or measurement matrix. Naturally, we wish that any pair of different sparse or approximately sparse signals could be mapped onto different projections under $A$, since we cannot reconstruct them from the same projection without additional information. Therefore, some properties such as the spark, the null space property (NSP) and the restricted isometry property (RIP) are introduced in the field of CS to characterize this capability of $A$ (a good reference about these properties is [9]).

In the theories of matrix, we can perform three types of elementary row operations on a matrix: row switching, row multiplication and row addition. Correspondingly, there are also three types of elementary column operations: column switching, column multiplication and column addition. These six types of transformations can be collectively referred to as elementary transformations. In this paper, we propose a question whether the spark, NSP and RIP of a sensing matrix will change after we apply elementary transformations to it. To the best of our knowledge, such a problem has not been previously addressed. And our research shows that all elementary transformations except column-addition ones do not change the spark, NSP order and RIP order of a sensing matrix. See Theorems 3.1-3.6 for details. Theorem 3.7 also shows that the highest order of NSP or RIP satisfied by a sensing matrix stays the same in these elementary transformations. Furthermore, Corollaries 3.1-3.3 tell us which types of matrices multiplying a sensing matrix can make the products preserve the three quantitative indexes mentioned above.

We denote three types of elementary matrices corresponding to three types of elementary row (or column) transformations respectively by: (1) $E_{ij}$ produced by exchanging row $i$ and row $j$ (or column $i$ and column $j$) of the identity matrix $I$. (2) $E_i(c)$ produced from $I$ by multiplying row $i$ (or column $i$) by $c$ ($c \neq 0$). (3) $E_{ij}(c)$ produced from $I$ by adding $c$ times row $i$ to row $j$ (or adding $c$ times column $j$ to column $i$).

The remainder of the paper is organized as follows. In Section 2, we will give some background of CS, especially the conditions under which exact recovery of all $k$-sparse signals is guaranteed. In Section 3, we will discuss the invariance of the spark, NSP order and RIP order under elementary transformations of matrices. Not only do we state our transformation invariance theorems and their corollaries in form of matrix products, but also analyze the influence of zero columns on the above quantitative indexes in column-addition transformations. In Section 4, we provide the proofs of our transformation invariance theorems. Finally, we conclude with an interesting result regarding the *universality* of sensing matrices and future research work.

## 2. Background

In this paper, the coherence [2, 10, 11, 17, 18] is not covered because it is not a quantitative index for exactly characterizing the ability of a sensing matrix to recover sparse signals although it can provide a bound for the spark. So we here introduce only the three properties in CS, i.e., the spark, NSP and RIP.

### 2.1 The spark

Obviously, if two distinct $k$-sparse signals $x$ and $x'$ sensed by a matrix $A$ have the same measurements $Ax = Ax' = y$, we can not distinguish or recover them based solely on $y$. So the sensing matrix $A$ must be an injective mapping for all vectors in $\Sigma_k$. If not, we will have $A(x - x') = \mathbf{0}$ and $x - x' \in \Sigma_{2k}$ in the case of $Ax = Ax'$. This implies that there exist some columns from A which are linearly dependent and whose number is not more than $2k$. Therefore, if the smallest number of columns from $A$ that are linearly dependent is greater than $2k$, we can guarantee that no pairs of distinct $k$-sparse signals have the same projection under $A$. This property of $A$ can be characterized by the spark [11] which is defined as follows.

**Definition 2.1.** *The spark of a given matrix A is the smallest number of columns of A that are linearly dependent.*

By definition, it is possible that more than one group consisting of spark columns of A is linearly dependent. Although the spark and the rank present some similarities in their definitions, there is no deterministic connection between them. For example, it is all possible for the spark to be less than, equal to, and more than the rank. It is not difficult to see that, for an $m \times n$ matrix $A$, its spark must be in the range $1 \leq \mathrm{spark}(A) \leq m + 1$.

The spark is one of the most common properties of sensing matrices used when recovering exactly sparse signals. It follows from above that $A$ can uniquely represents all $\boldsymbol{x} \in \Sigma_k$ if the spark of $A$ is greater than $2k$. And it can be proved that "$\mathrm{spark}(A) > 2k$" is the necessary and sufficient condition for exact recovery of all $k$-sparse signals [9, 11].

**2.2 The null space property**

When dealing with approximately $k$-sparse signals, if two distinct signals $\boldsymbol{x}$ and $\boldsymbol{x}'$ have the same projection under $A$ then we have $A(\boldsymbol{x} - \boldsymbol{x}') = \boldsymbol{0}$ and $\boldsymbol{x} - \boldsymbol{x}'$ can be well-approximated by a $2k$-sparse vector. So we need to use the null space property (NSP) [8] to ensure that the null space of $A$, denoted by $\mathcal{N}(A)$, does not contain any approximately $2k$-sparse vectors in addition to vectors in $\Sigma_{2k}$.

**Definition 2.2.** *A matrix $A$ satisfies the NSP of order $k$ if there exists a constant $C > 0$ such that*

$$\|\boldsymbol{h}_\Lambda\|_2 \leq C \frac{\|\boldsymbol{h}_{\Lambda^c}\|_1}{\sqrt{k}} \tag{2.1}$$

*holds for all $\boldsymbol{h} \in \mathcal{N}(A)$ and for all $\Lambda$ such that $|\Lambda| \leq k$.*

By definition, if $A$ satisfies the NSP of order $k$ ($k \geq 1$) then it necessarily satisfies the NSP of any order smaller than $k$, which makes the NSP order has downward compatibility. On the other hand, if $A$ does not satisfy the NSP of order $k$ then it certainly does not satisfy the NSP of any order larger than $k$, which makes the NSP order have an upper bound, i.e., the highest order. Since order $k$ of the NSP satisfied by sensing matrices is not necessarily the highest order, it could be any value from 1 to the highest order.

When (2.1) holds, together with Lemma 1.2 of [9], we can infer that $\|\boldsymbol{h}_\Lambda\|_1 \leq \sqrt{k}\|\boldsymbol{h}_\Lambda\|_2 \leq C\|\boldsymbol{h}_{\Lambda^c}\|_1 < C\|\boldsymbol{h}\|_1$. Then let $C$ be a value not greater than 1, we thereby have some definitions of the NSP [13-15] which can be derived from (2.1). For this reason, we select Definition 2.2 as the NSP definition while discussing invariance of the NSP order under elementary transformations.

The significance of the NSP lies in that it can provide a performance guarantee for sparse recovery algorithms in the form of

$$\|\hat{x} - x\|_2 \leq C \frac{\sigma_k(x)_1}{\sqrt{k}}, \tag{2.2}$$

where $\sigma_k(x)_1$ represents the error of the best $k$-term approximation to a sparse or non-sparse signal $x$ (approximation by the nearest $k$-sparse vector to $x$), and $\hat{x}$ is a vector found by the algorithms subject to $A\hat{x} = Ax$. It is easy to see that any recovery algorithm satisfying (2.2) can guarantee, in noiseless settings, the exact recovery of all $k$-sparse signals $x$ and an upper bound for the recovery error of non-sparse signals $x$ that depends on how well $x$ is approximated by a $k$-sparse vector.

Theorems in [8, 9, 19] have shown that "$A$ satisfies the NSP of order $2k$ (or above)" is a necessary and sufficient condition for $\ell_1$-minimization recovery algorithms to satisfy (2.2). Therefore, the NSP order is a quantitative index which can be used to guarantee exact signal recovery in noiseless settings.

**2.3 The restricted isometry property**

When dealing with measurements contaminated with noise, we need to use properties in measurement form to account for noise and, at the same time, to provide a performance guarantee for sparse recovery algorithms. To this end, the restricted isometry property (RIP) [4] is introduced into the domain of CS.

**Definition 2.3.** *A matrix A satisfies the RIP of order k if there exists a $\delta_k \in (0, 1)$ such that*

$$(1-\delta_k)\|x\|_2^2 \le \|Ax\|_2^2 \le (1+\delta_k)\|x\|_2^2 \qquad (2.3)$$

*holds for all $x \in \Sigma_k$.*

As a matter of fact, the bounds in the RIP definition need not be symmetric about 1. An asymmetric version of the RIP [9] is defined as the following.

**Definition 2.4.** *A matrix A satisfies the RIP of order k if there exist constants $\alpha$ and $\beta$ ($0 < \alpha \le \beta < \infty$) such that*

$$\alpha\|x\|_2^2 \le \|Ax\|_2^2 \le \beta\|x\|_2^2 \qquad (2.4)$$

*holds for all $x \in \Sigma_k$.*

For convenience, we will assume the matrix $A$ satisfies the asymmetric RIP in proving transformation invariance theorems of the RIP order. The reason for doing so is that we can easily convert an asymmetric version of the RIP with constants $\alpha$ and $\beta$ into a symmetric version of the RIP with a constant $\delta_k = (\beta - \alpha)/(\beta + \alpha)$ if we multiply two sides of the inequalities in (2.4) by $2/(\beta + \alpha)$.

From the RIP definition, we can see that the RIP has downward compatibility with respect to orders and the highest order as well as the NSP. That is to say, if $A$ satisfies the RIP of order $k$ ($k \ge 1$) then it necessarily satisfies the RIP of any order less than $k$, and if $A$ does not satisfy the RIP of order $k$ then it certainly does not satisfy the RIP of any order more than $k$. If order $k$ of the RIP satisfied by sensing matrices is not specified as the highest order, then it could be any value from 1 to the highest order.

The RIP of order $2k$ ensures that the distance between any pair of $k$-sparse signals is approximately preserved in their projections under the mapping $A$, which brings to measurements of the signals robustness to noise. And the measurement form of the RIP makes it more applicable to noisy measurements than the NSP.

It has been proved that the NSP is included in the RIP, and "$A$ satisfies the RIP of order $2k$ (or above)" is a sufficient condition for $\ell_1$-minimization recovery algorithms to satisfy (2.2) [3, 6, 9]. Therefore, the RIP order is also a quantitative index used to guarantee exact signal recovery in noise-free settings.

**3. Transformation invariance of the spark, NSP order and RIP order**

Based on our research, we point out that all elementary transformations except column-addition ones can keep the spark, NSP order and RIP order from changing. And we will exemplify that, in column-addition transformations, the appearing and disappearing of zero columns is an important cause of the break of transformation invariance.

**3.1 Transformation invariance theorems**

Here we propose our theorems to describe invariance of the spark, NSP order and RIP order under elementary transformations of matrices. For convenience, the theorems are referred to as transformation invariance theorems.

Theorems 3.1-3.6 below give the types of elementary transformations under which values of the spark, orders of the NSP and RIP remain the same. We will prove these theorems in Section 4.

**Theorem 3.1.** *If we transform a matrix A into B by executing elementary row operations, then we have* $\mathrm{spark}(B) = \mathrm{spark}(A)$.

**Theorem 3.2.** *If we transform a matrix A into B by executing elementary column operations except column addition, then we have* $\mathrm{spark}(B) = \mathrm{spark}(A)$.

**Theorem 3.3.** *Assume that a matrix A satisfies the NSP of order k. If we transform A into B by executing elementary row operations, then B still satisfies the NSP of order k.*

**Theorem 3.4.** *Assume that a matrix A satisfies the NSP of order k. If we transform A into B by executing elementary column operations except column addition, then B still satisfies the NSP of order k.*

**Theorem 3.5.** *Assume that a matrix A satisfies the RIP of order k. If we transform A into B by executing elementary row operations, then B still satisfies the RIP of order k.*

**Theorem 3.6.** *Assume that a matrix A satisfies the RIP of order k. If we transform A into B by executing elementary column operations except column addition, then B still satisfies the RIP of order k.*

We already know that the NSP or RIP satisfied by sensing matrices has the highest order. In Theorems 3.3-3.6, we do not suppose that $k$ is the highest order of the NSP or RIP satisfied by $A$. So we have the following question: if $k$ is the highest order of the NSP or RIP satisfied by $A$, then is it also the highest order of the NSP or RIP satisfied by $B$? In other words, is the highest order of the NSP or RIP also unchanged in those elementary transformations which can keep their orders from changing? Based on Theorem 3.7 below, our answer is affirmative.

**Theorem 3.7.** *Assume that the highest order of the NSP or RIP satisfied by a matrix A is k. If we transform A into B by executing elementary row or column operations except column addition, then k is also the highest order of the NSP or RIP satisfied by B.*

*Proof.* We only need to prove that every time when we apply one of the five types of operations, i.e., row or column switching, row or column multiplication and row addition, to the matrix $A$, the highest order of the NSP or RIP satisfied by $B$ is also $k$.

We first know from Theorems 3.3-3.6 that $B$ satisfies the NSP or RIP of order $k$. Now suppose for the sake of a contradiction that $B$ satisfies the NSP or RIP of order $k+1$ or above. We have the following facts: (1) If we get $B$ by performing an operation of row or column switching on $A$, which equals to $E_{ij}A$ or $AE_{ij}$, then we can obtain $A$ by performing an operation of the same type on $B$, which equals to $E_{ij}B$ or $BE_{ij}$. (2) If we get $B$ by performing an operation of row or column multiplication on $A$, which equals to $E_i(c)A$ or $AE_i(c)$, then we can obtain $A$ by performing an operation of the same type on $B$, which equals to $E_i(1/c)B$ or $BE_i(1/c)$. (3) If we get $B$ by performing an operation of row addition on $A$, which equals to $E_{ij}(c)A$, then we can obtain $A$ by performing an operation of the same type on $B$, which equals to $E_{ij}(-c)B$. For the above operations used to transform $B$ into $A$, Theorems 3.3-3.6 together with our assumption tell us that these operations make $A$ satisfy the NSP or RIP of the same order as $B$, i.e., $k+1$ or above. This contradicts the assumption of this theorem. Whence, $k$ is also the highest order of the NSP or RIP satisfied by $B$. □

Grounded on Theorems 3.3-3.7, we arrive at a conclusion that not only orders but also the highest order of the NSP or RIP satisfied by a sensing matrix does not change under all elementary transformations except column-addition ones. From the proof of Theorem 3.7, we can see that if some kind of elementary row or column operations can keep orders of the NSP or RIP unchanged, then it can also keep the highest order of the NSP or RIP unchanged. The converse is obviously true. Thus, any type of elementary transformations preserves the highest order of the NSP or RIP if and only if it preserves orders of the NSP or RIP.

We know that applying an elementary row or column operation to a matrix $A$ is equivalent to multiplying $A$ by its corresponding elementary matrix on the left or right. The product of any combination of three types of elementary matrices, i.e., $E_{ij}$, $E_i(c)$ and $E_{ij}(c)$, is an invertible matrix, and the product of any combination of $E_{ij}$ and $E_i(c)$ is an invertible diagonal matrix or a matrix produced by any permutation of its rows or columns. Therefore, we get the following corollaries if we describe the above Theorems 3.1-3.6 in form of matrix products.

**Corollary 3.1.** *Let $B_1$ be an invertible matrix, and $B_2$ be a matrix produced by any permutation of the rows or columns of an invertible diagonal matrix. For a matrix A, we have* $\text{spark}(B_1A) = \text{spark}(A)$ *and* $\text{spark}(AB_2) = \text{spark}(A)$.

**Corollary 3.2.** *Let $B_1$ be an invertible matrix, and $B_2$ be a matrix produced by any permutation of the rows or columns of an invertible diagonal matrix. If a matrix A satisfies the NSP of order k, then $B_1A$ and $AB_2$ still satisfy the NSP of order k.*

**Corollary 3.3.** *Let $B_1$ be an invertible matrix, and $B_2$ be a matrix produced by any permutation of the rows or columns of an invertible diagonal matrix. If a matrix A satisfies the RIP of order k, then $B_1A$ and $AB_2$ still satisfy the RIP of order k.*

**3.2 Influence of zero columns on transformation invariance**

All of the spark, NSP order and RIP order do not possess invariance under column-addition transformations because we can easily give corresponding counter examples. In these counter examples, we find out that the appearing and disappearing of zero columns as a result of these transformations is an important reason for the break of invariance. Since column-addition transformations do not preserve orders of the NSP or RIP, this type of elementary transformations can not preserve the highest order of the NSP or RIP. For example, suppose that a matrix $A$ satisfies the NSP or RIP of order $k$ and the highest order is also $k$. After we get $B$ by doing column-addition transformations on $A$, it is possible for the highest order to become smaller or bigger. Clearly, $B$ does not satisfy the NSP or RIP of order $k$ in the former case, whereas it still does in the latter case. Here we make some discussion only about the influence of zero columns on invariance of the spark, NSP order and RIP order under column-addition transformations.

### 3.2.1 Relationship between zero columns and the spark

A matrix $A$ contains zero columns if and only if spark$(A) = 1$, because a single zero vector is linearly dependent whereas a single nonzero vector is not. Thus, if we perform elementary transformations on $A$ and spark$(A) \geq 2$ is not specified, we must consider the case where $A$ contains zero columns before the transformations. And we can easily find two examples related to zero columns which are used to disprove invariance of the spark under column-addition transformations. One of the examples is that a matrix $A$ contains a zero column and the only zero column disappears after applying an operation of column addition to $A$, which means the spark changes from 1 to 2. The other example is that $A$ contains no zero columns and a zero column appears after applying an operation of column addition to $A$, which means the spark changes from 2 to 1.

We know that, for a sensing matrix $A$ without zero columns, because some or all columns of $A$ are linearly dependent, zero columns will be sure to appear after doing enough times of column-addition transformations on $A$. If $A$ contains two columns that are linearly dependent, we can get a zero column by performing an appropriate operation of column addition between the two columns. If at least three columns of $A$ are needed to become linearly dependent, we can get a zero column by performing at least two appropriate operations of column addition among these columns. In the same way, by definition, at least spark columns of $A$ are needed to become linearly dependent, therefore we can not get a zero column until we perform at least spark$-1$ times of column-addition transformations on $A$.

### 3.2.2 Relationship between zero columns and the NSP or RIP order

A matrix $A$ will not satisfy the NSP or RIP of any order when zero columns emerge in $A$ as a consequence of column-addition transformations. The two propositions below can prove the claim.

**Proposition 3.1.** *If a matrix A contains zero columns, then A does not satisfy the NSP of order* $k \geq 1$.

*Proof.* Let $i$ be the column number of one of zero columns of $A$. There exists a vector $\boldsymbol{h} = (0, \ldots, h_i, \ldots, 0)^T$ (its only nonzero entry is $h_i \neq 0$) such that $A\boldsymbol{h} = \boldsymbol{0}$, i.e., $\boldsymbol{h} \in \mathcal{N}(A)$. Also there exists a $\Lambda$ such that $i \in \Lambda$ and $|\Lambda| = k$ ($k \geq 1$). Then we have $\|\boldsymbol{h}_\Lambda\|_2 = \sqrt{h_i^2} > 0$. On the other hand, we know from $i \notin \Lambda^c$ that $\|\boldsymbol{h}_{\Lambda^c}\|_1 = 0$. Hence, according to (2.1), $A$ does not satisfy the NSP of order $k \geq 1$. □

**Proposition 3.2.** *If a matrix A contains zero columns, then A does not satisfy the RIP of order $k \geq 1$.*

*Proof.* Let $i$ be the column number of one of zero columns of $A$. There exists a $k$-sparse vector ($k \geq 1$) $\boldsymbol{x} = (0, \ldots, x_i, \ldots, 0)^T$ (its only nonzero entry is $x_i \neq 0$) such that $\|A\boldsymbol{x}\|_2^2 = 0$ and $\|\boldsymbol{x}\|_2^2 = x_i^2 > 0$. Thus, according to (2.3), $A$ does not satisfy the RIP of order $k \geq 1$. □

From the two propositions above, we can derive that if we perform elementary transformations on a matrix $A$ and $A$ satisfies the NSP or RIP of order $k \geq 1$, we need not consider the case where $A$ has zero columns before the transformations. And we can also get an example concerning zero columns which are used to disprove invariance of the NSP or RIP order under column-addition transformations. If $A$ satisfies the NSP or RIP of order $k$ and a zero column emerges after performing an operation of column addition on $A$, then the order of the NSP or RIP changes from $k$ to nothing.

When a matrix $A$ satisfies the NSP or RIP of order $k$, it is easy to see, by definition, that $\mathcal{N}(A)$ contains no vectors in $\Sigma_k$. This implies the number of columns of $A$ that are linearly dependent is more than $k$ (the possible minimum is $k + 1$). Therefore, we can not get a zero column until we perform at least $k$ times of column-addition transformations on $A$.

## 4. Proofs of transformation invariance theorems

Below we give the proofs of transformation invariance theorems of the spark, NSP order and RIP order, i.e., Theorems 3.1-3.6.

### 4.1 Proofs of Theorems 3.1-3.2

To begin with, we deduce several helpful facts to be used in the proofs from the spark definition. For a matrix $A$, we have: (1) Any group of columns of $A$ whose number is smaller than spark($A$) is necessarily linearly independent. (2) If a group of columns of $A$ is linearly dependent, then its number of columns is certainly bigger than or equal to spark($A$). The converse is not necessarily true. (3) Given a number $r \geq$ spark($A$), there always exists a group of $r$ columns of $A$ which is linearly dependent, since we can get this group by adding $r -$ spark($A$) columns to some group which is made up of spark($A$) columns of $A$ that are linearly dependent.

Next, we need a lemma as follows in order to prove Theorem 3.1.

**Lemma 4.1.** *If we transform a matrix A into B by performing elementary row operations, then any group of columns from A and its counterpart from B have the same linear dependence, i.e., if*

$$A = (\boldsymbol{\alpha}_1, \boldsymbol{\alpha}_2, \ldots, \boldsymbol{\alpha}_n) \xrightarrow{elementary\ row\ operations} (\boldsymbol{\beta}_1, \boldsymbol{\beta}_2, \ldots, \boldsymbol{\beta}_n) = B,$$

*then two groups of columns, $\boldsymbol{\alpha}_{i_1}, \boldsymbol{\alpha}_{i_2}, \ldots, \boldsymbol{\alpha}_{i_r}$ and $\boldsymbol{\beta}_{i_1}, \boldsymbol{\beta}_{i_2}, \ldots, \boldsymbol{\beta}_{i_r}$ ($1 \leq i_1 < i_2 < \cdots < i_r \leq n$), have the same linear dependence.*

*Proof.* To transform $A$ into $B$ by doing elementary row operations, one can multiply $A$ by a number of elementary matrices $P_1, \ldots, P_s$ on the left and make the product equal to $B$. If we denote $P_s \cdots P_2 P_1$ by $P$, then we have $PA = B$ and hence $P\alpha_j = \beta_j$, $j = 1, 2, \ldots, n$. Let $A_I = (\alpha_{i_1}, \alpha_{i_2}, \ldots, \alpha_{i_r})$, $B_I = (\beta_{i_1}, \beta_{i_2}, \ldots, \beta_{i_r})$ and $x_I = (x_{i_1}, x_{i_2}, \ldots, x_{i_r})^T$. Obviously, two homogeneous systems of linear equations $A_I x_I = 0$ and $B_I x_I = 0$ (i.e., $PA_I x_I = 0$) have the same solutions. And the columns of $A_I$ or $B_I$ are linearly dependent if and only if their corresponding systems of equations have nonzero solutions. Therefore, the columns of $A_I$ and $B_I$ have the same linear dependence. □

Then, we give the proofs of Theorem 3.1 and Theorem 3.2.

**Theorem 3.1.** *If we transform a matrix A into B by executing elementary row operations, then we have* $\text{spark}(B) = \text{spark}(A)$.

*Proof.* Let $A$ be an $m \times n$ matrix and its $n$ columns are denoted $\alpha_1, \alpha_2, \ldots, \alpha_n$. We get $B$ by executing elementary row operations on $A$ and denote its $n$ columns by $\beta_1, \beta_2, \ldots, \beta_n$. We consider two cases below.

(a) Assume that $\text{spark}(B) \leq r < \text{spark}(A)$. Because $r \geq \text{spark}(B)$, there exist necessarily $r$ columns of $B$ that are linearly dependent. Let $\beta_{i_1}, \beta_{i_2}, \ldots, \beta_{i_r}$ and $\alpha_{i_1}, \alpha_{i_2}, \ldots, \alpha_{i_r}$ ($1 \leq i_1 < i_2 < \cdots < i_r \leq n$) be the $r$ columns of $B$ and their counterparts of $A$ respectively. By Lemma 4.1, the $r$ columns $\alpha_{i_1}, \alpha_{i_2}, \ldots, \alpha_{i_r}$ of $A$ are also linearly dependent. This implies $r \geq \text{spark}(A)$, which contradicts the assumption. Hence, we have $\text{spark}(B) \geq \text{spark}(A)$.

(b) Assume that $\text{spark}(B) > r \geq \text{spark}(A)$. In the same way as (a), we can arrive at a contradiction with the assumption. So we get $\text{spark}(B) \leq \text{spark}(A)$.

Combining (a) with (b), we get that $\text{spark}(B) = \text{spark}(A)$. □

**Theorem 3.2.** *If we transform a matrix A into B by executing elementary column operations except column addition, then we have* $\text{spark}(B) = \text{spark}(A)$.

*Proof.* We only need to prove that every time when we perform an operation of column switching or column multiplication, the spark of matrices remains the same. Let $A$ be an $m \times n$ matrix and its $n$ columns are denoted $\alpha_1, \alpha_2, \ldots, \alpha_n$.

(i) By exchanging the positions of two columns of $A$, we get the matrix $B$ whose $n$ columns are still the $n$ columns of $A$. Of course, $\text{spark}(B) = \text{spark}(A)$.

(ii) By multiplying column $i$ of $A$ by a nonzero constant $c$, we get $B$ whose $n$ columns are $\alpha_1, \alpha_2, \ldots, c\alpha_i, \ldots, \alpha_n$. We know that when $A$ contains (no) zero columns, so does (not) $B$ after the column multiplication. When $A$ has zero columns, we have $\text{spark}(B) = \text{spark}(A) = 1$. When $A$ possesses no zero columns, we may consider two assumptions of $\text{spark}(B) \neq \text{spark}(A)$ below.

(a) Assume that $\text{spark}(B) \leq r < \text{spark}(A)$. Because $r \geq \text{spark}(B)$, there exist necessarily $r$ columns of $B$ that are linearly dependent. If the $r$ columns of $B$ do not contain $c\alpha_i$, they are also $r$ columns of $A$ that are linearly dependent and thus we get $r \geq \text{spark}(A)$, which contradicts the assumption. Therefore, we may denote the $r$ columns of $B$ by $c\phi_1, \phi_2, \ldots, \phi_r$ (where $\phi_1 = \alpha_i$). And we can find $k_1, k_2, \ldots, k_r \in R$ not all zero so that $k_1 c\phi_1 + k_2 \phi_2 + \cdots + k_r \phi_r = \mathbf{0}$. If $k_1 \neq 0$, then by combining it with $c \neq 0$, we know that $A$ has $r$ columns $\phi_1, \phi_2, \ldots, \phi_r$ that are linearly dependent and hence we have $r \geq \text{spark}(A)$, which contradicts the assumption. If $k_1 = 0$, then we know that $A$ has $r - 1$ columns $\phi_2, \phi_3, \ldots, \phi_r$ that are linearly dependent and thus we get $r - 1 \geq \text{spark}(A)$, which also contradicts the assumption. Based on the above contradictions, we have $\text{spark}(B) \geq \text{spark}(A)$.

(b) Assume that $\text{spark}(B) > r \geq \text{spark}(A)$. Because $r \geq \text{spark}(A)$, there exist necessarily $r$ columns of $A$ that are linearly dependent. If the $r$ columns of $A$ do not contain the $i$ th column of $A$, i.e., $\alpha_i$, they are also $r$ columns of $B$ that are linearly dependent and then we have $r \geq \text{spark}(B)$, which contradicts the assumption. Therefore, we may denote the $r$ columns of $A$ by $\phi_1, \phi_2, \ldots, \phi_r$ (where $\phi_1 = \alpha_i$). And we can find $k_1, k_2, \ldots, k_r \in R$ not all zero so that $k_1 \phi_1 + k_2 \phi_2 + \cdots + k_r \phi_r = \mathbf{0}$. Since $c \neq 0$, there exist $k_1/c, k_2, \ldots, k_r \in R$ not all zero so that $(k_1/c)(c\phi_1) + k_2 \phi_2 + \cdots + k_r \phi_r = \mathbf{0}$. This shows $B$ has $r$ columns $c\phi_1, \phi_2, \ldots, \phi_r$ that are linearly dependent and then we get $r \geq \text{spark}(B)$, which contradicts the assumption. Thus, we have $\text{spark}(B) \leq \text{spark}(A)$.

Combining (a) with (b), we know that when $A$ has no zero columns, we also have $\text{spark}(B) = \text{spark}(A)$. So this is the end of (ii).

Finally, this theorem is proven by (i) and (ii). □

**4.2 Proofs of Theorems 3.3-3.4**

**Theorem 3.3.** *Assume that a matrix $A$ satisfies the NSP of order $k$. If we transform $A$ into $B$ by executing elementary row operations, then $B$ still satisfies the NSP of order $k$.*

*Proof.* To apply elementary row operations to $A$, one can multiply $A$ by their corresponding elementary matrices on the left. So we denote the product of these elementary matrices by $P$ and then we have $PA = B$. According to Gaussian elimination for solving systems of linear equations, $PA\mathbf{h} = \mathbf{0}$ and $A\mathbf{h} = \mathbf{0}$ have the same solutions. Hence, for all $\mathbf{h} \in \mathcal{N}(PA)$, we have $\mathbf{h} \in \mathcal{N}(A)$. Because $A$ satisfies the NSP of order $k$, so does $PA$ based on the NSP definition. This proves the theorem. □

**Theorem 3.4.** *Assume that a matrix $A$ satisfies the NSP of order $k$. If we transform $A$ into $B$ by executing elementary column operations except column addition, then $B$ still satisfies the NSP of order $k$.*

*Proof.* We only need to prove that every time when we do an operation of column switching or column multiplication, the NSP order of matrices remains the same. Let $A$ be an $m \times n$ matrix and its $n$ columns are denoted $\alpha_1, \alpha_2, \ldots, \alpha_n$. Since $A$ satisfies the NSP of order $k$, $A$ contains no zero columns and so does the matrix obtained after the operation of column switching or column multiplication.

(i) To swap the positions of two columns of $A$, one can multiply $A$ by the elementary matrix $E_{ij}$ on the right. For any $\mathbf{h} \in \mathcal{N}(AE_{ij})$, we have

$$AE_{ij}\mathbf{h} = AE_{ij}(h_1, \ldots, h_i, \ldots, h_j, \ldots, h_n)^T = A(h_1, \ldots, h_j, \ldots, h_i, \ldots, h_n)^T = \mathbf{0},$$

and we get $E_{ij}\mathbf{h} \in \mathcal{N}(A)$. Obviously, $E_{ij}\mathbf{h}$ merely changes the places of two coefficients of $\mathbf{h}$ but not their magnitude. Thus, for all $\Lambda$ such that $|\Lambda| \leq k$, $\mathbf{h}$ satisfies (2.1) as long as $E_{ij}\mathbf{h}$ does. Since $A$ satisfies the NSP of order $k$, so does $AE_{ij}$.

(ii) To multiply the $i$ th column of $A$ by a nonzero constant $c$, one can multiply $A$ by the elementary matrix $E_i(c)$ ($c \neq 0$) on the right. For any $\boldsymbol{h} \in \mathcal{N}(AE_i(c))$, we have
$$AE_i(c)\boldsymbol{h} = AE_i(c)(h_1, \ldots, h_i, \ldots, h_n)^T = A(h_1, \ldots, ch_i, \ldots, h_n)^T = \boldsymbol{0},$$
and we get $E_i(c)\boldsymbol{h} \in \mathcal{N}(A)$.

For any $\Lambda$ such that $|\Lambda| = K \leq k$, we consider two situations where $i \in \Lambda$ and $i \notin \Lambda$.

(a) When $i \in \Lambda$, let, without loss of generality, $\Lambda = \{l_0, l_1, \ldots, l_{K-1}\}$ and $l_0 = i$. Since $i \notin \Lambda^c$, we have $\|\boldsymbol{h}_{\Lambda^c}\|_1 = \|(E_i(c)\boldsymbol{h})_{\Lambda^c}\|_1$.

If $0 < |c| \leq 1$, then we have
$$|c| \cdot \|\boldsymbol{h}_\Lambda\|_2 = \sqrt{c^2 h_i^2 + c^2 h_{l_1}^2 + \cdots + c^2 h_{l_{K-1}}^2} \leq \sqrt{(ch_i)^2 + h_{l_1}^2 + \cdots + h_{l_{K-1}}^2} = \|(E_i(c)\boldsymbol{h})_\Lambda\|_2.$$

Thus we get
$$\|\boldsymbol{h}_\Lambda\|_2 \leq \frac{1}{|c|} \cdot \|(E_i(c)\boldsymbol{h})_\Lambda\|_2 \leq \frac{1}{|c|} \cdot C \frac{\|(E_i(c)\boldsymbol{h})_{\Lambda^c}\|_1}{\sqrt{k}} = \frac{1}{|c|} C \frac{\|\boldsymbol{h}_{\Lambda^c}\|_1}{\sqrt{k}},$$
where the second inequality follows from $E_i(c)\boldsymbol{h} \in \mathcal{N}(A)$ and that $A$ satisfies the NSP of order $k$.

If $|c| > 1$, then we have
$$\|\boldsymbol{h}_\Lambda\|_2 = \sqrt{h_i^2 + h_{l_1}^2 + \cdots + h_{l_{K-1}}^2} < \sqrt{(ch_i)^2 + h_{l_1}^2 + \cdots + h_{l_{K-1}}^2} = \|(E_i(c)\boldsymbol{h})_\Lambda\|_2.$$

So we obtain
$$\|\boldsymbol{h}_\Lambda\|_2 < \|(E_i(c)\boldsymbol{h})_\Lambda\|_2 \leq C \frac{\|(E_i(c)\boldsymbol{h})_{\Lambda^c}\|_1}{\sqrt{k}} = C \frac{\|\boldsymbol{h}_{\Lambda^c}\|_1}{\sqrt{k}},$$
where the second inequality follows from $E_i(c)\boldsymbol{h} \in \mathcal{N}(A)$ and that $A$ satisfies the NSP of order $k$.

(b) When $i \notin \Lambda$, we have $\|\boldsymbol{h}_\Lambda\|_2 = \|(E_i(c)\boldsymbol{h})_\Lambda\|_2$. Since $i \in \Lambda^c$, let, without loss of generality, $\Lambda^c = \{l_0, l_1, \ldots, l_{n-K-1}\}$ and $l_0 = i$.

If $0 < |c| \leq 1$, then we have
$$\|(E_i(c)\boldsymbol{h})_{\Lambda^c}\|_1 = |ch_i| + |h_{l_1}| + \cdots + |h_{l_{n-K-1}}| \leq |h_i| + |h_{l_1}| + \cdots + |h_{l_{n-K-1}}| = \|\boldsymbol{h}_{\Lambda^c}\|_1.$$

So we get
$$\|\boldsymbol{h}_\Lambda\|_2 = \|(E_i(c)\boldsymbol{h})_\Lambda\|_2 \leq C \frac{\|(E_i(c)\boldsymbol{h})_{\Lambda^c}\|_1}{\sqrt{k}} \leq C \frac{\|\boldsymbol{h}_{\Lambda^c}\|_1}{\sqrt{k}},$$
where the first inequality follows from $E_i(c)\boldsymbol{h} \in \mathcal{N}(A)$ and that $A$ satisfies the NSP of order $k$.

If $|c| > 1$, then we have
$$\frac{1}{|c|} \cdot \|(E_i(c)\boldsymbol{h})_{\Lambda^c}\|_1 = \frac{1}{|c|} \cdot |ch_i| + \frac{1}{|c|} \cdot |h_{l_1}| + \cdots + \frac{1}{|c|} \cdot |h_{l_{n-K-1}}| < |h_i| + |h_{l_1}| + \cdots + |h_{l_{n-K-1}}| = \|\boldsymbol{h}_{\Lambda^c}\|_1.$$

Thus we obtain
$$\|\boldsymbol{h}_\Lambda\|_2 = \|(E_i(c)\boldsymbol{h})_\Lambda\|_2 \leq C \frac{\|(E_i(c)\boldsymbol{h})_{\Lambda^c}\|_1}{\sqrt{k}} < |c| \cdot C \frac{\|\boldsymbol{h}_{\Lambda^c}\|_1}{\sqrt{k}},$$

where the first inequality follows from $E_i(c)\boldsymbol{h} \in \mathcal{N}(A)$ and that $A$ satisfies the NSP of order $k$.

Combining (a) with (b), we can see that let $C_0 = \max\{|c|\cdot C, C/|c|\}$, then $\|\boldsymbol{h}_\Lambda\|_2 \leq C_0 \dfrac{\|\boldsymbol{h}_{\Lambda^c}\|_1}{\sqrt{k}}$

holds for all $\boldsymbol{h} \in \mathcal{N}(AE_i(c))$ and for all $\Lambda$ such that $|\Lambda| \leq k$. This shows $AE_i(c)$ satisfies the NSP of order $k$. $\square$

### 4.3 Proofs of Theorems 3.5-3.6

**Theorem 3.5.** *Assume that a matrix $A$ satisfies the RIP of order $k$. If we transform $A$ into $B$ by executing elementary row operations, then $B$ still satisfies the RIP of order $k$.*

*Proof.* We only need to prove that every time when we do an operation of row switching, row multiplication and row addition, the RIP order of matrices remains the same. Let $A$ be an $m \times n$ matrix and assume that $A$ satisfies the asymmetric RIP of order $k$.

(i) To exchange the positions of two rows of $A$, one can multiply $A$ by the elementary matrix $E_{ij}$ on the left. For any $\boldsymbol{x} \in \Sigma_k$, if we let $A\boldsymbol{x} = \boldsymbol{y}$, then we can get
$$E_{ij}A\boldsymbol{x} = E_{ij}\boldsymbol{y} = E_{ij}(y_1, \ldots, y_i, \ldots, y_j, \ldots, y_m)^T = (y_1, \ldots, y_j, \ldots, y_i, \ldots, y_m)^T.$$

So we can easily see that $\|E_{ij}A\boldsymbol{x}\|_2^2 = \|A\boldsymbol{x}\|_2^2$. Because $A$ satisfies (2.4) according to the assumption, we obtain that $\alpha\|\boldsymbol{x}\|_2^2 \leq \|E_{ij}A\boldsymbol{x}\|_2^2 \leq \beta\|\boldsymbol{x}\|_2^2$, which shows that $E_{ij}A$ satisfies the asymmetric RIP of order $k$.

(ii) To multiply the $i$ th row of $A$ by a nonzero constant $c$, one can multiply $A$ by the elementary matrix $E_i(c)$ ($c \neq 0$) on the left. For any $\boldsymbol{x} \in \Sigma_k$, if we let $A\boldsymbol{x} = \boldsymbol{y}$, then we can get
$$E_i(c)A\boldsymbol{x} = E_i(c)\boldsymbol{y} = E_i(c)(y_1, \ldots, y_i, \ldots, y_m)^T = (y_1, \ldots, cy_i, \ldots, y_m)^T.$$
Thus we have
$$\|E_i(c)A\boldsymbol{x}\|_2^2 = y_1^2 + \cdots + (cy_i)^2 + \cdots + y_m^2. \tag{4.1}$$

Due to the value of $c$, we need to consider two situations below.

If $0 < |c| \leq 1$, then it follows from (4.1) that
$$\|E_i(c)A\boldsymbol{x}\|_2^2 \leq y_1^2 + \cdots + y_i^2 + \cdots + y_m^2 = \|A\boldsymbol{x}\|_2^2,$$
and
$$\|E_i(c)A\boldsymbol{x}\|_2^2 \geq c^2 y_1^2 + \cdots + c^2 y_i^2 + \cdots + c^2 y_m^2 = c^2\|A\boldsymbol{x}\|_2^2.$$

These two inequalities, together with the assumption that $A$ satisfies (2.4), tell us that $c^2\alpha\|\boldsymbol{x}\|_2^2 \leq \|E_i(c)A\boldsymbol{x}\|_2^2 \leq \beta\|\boldsymbol{x}\|_2^2$.

If $|c| > 1$, then it follows from (4.1) that
$$\|E_i(c)A\boldsymbol{x}\|_2^2 > y_1^2 + \cdots + y_i^2 + \cdots + y_m^2 = \|A\boldsymbol{x}\|_2^2,$$
and
$$\|E_i(c)A\boldsymbol{x}\|_2^2 < c^2 y_1^2 + \cdots + c^2 y_i^2 + \cdots + c^2 y_m^2 = c^2\|A\boldsymbol{x}\|_2^2.$$

These two inequalities, together with the assumption that $A$ satisfies (2.4), give us that $\alpha\|x\|_2^2 \leq \|E_i(c)Ax\|_2^2 \leq c^2\beta\|x\|_2^2$.

From the above two situations, we can see that let $\alpha_0 = \min\{\alpha, c^2\alpha\}$ and $\beta_0 = \max\{\beta, c^2\beta\}$, then $\alpha_0\|x\|_2^2 \leq \|E_i(c)Ax\|_2^2 \leq \beta_0\|x\|_2^2$ holds for all $x \in \Sigma_k$. This indicates $E_i(c)A$ satisfies the asymmetric RIP of order $k$.

(iii) To multiply the $i$ th row of $A$ by a constant $c$ and then add it to the $j$ th row of $A$, one can multiply $A$ by the elementary matrix $E_{ij}(c)$ on the left. For any $x \in \Sigma_k$, if we let $Ax = y$, then we can get

$E_{ij}(c)Ax = E_{ij}(c)y = E_{ij}(c)(y_1, \ldots, y_i, \ldots, y_j, \ldots, y_m)^T = (y_1, \ldots, y_i, \ldots, cy_i + y_j, \ldots, y_m)^T$.

Thus we have

$$\|E_{ij}(c)Ax\|_2^2 = y_1^2 + \cdots + y_i^2 + \cdots + (cy_i + y_j)^2 + \cdots + y_m^2. \tag{4.2}$$

The only difference between $\|E_{ij}(c)Ax\|_2^2$ and $\|Ax\|_2^2$ is their $j$ th entries, so we get

$$\|E_{ij}(c)Ax\|_2^2 - (cy_i + y_j)^2 = \|Ax\|_2^2 - y_j^2. \tag{4.3}$$

On the other hand, we have that $\|x\|_2^2 \neq 0$ for all $x \in \Sigma_k$. Hence, from the assumption that $A$ satisfies (2.4), we know that

$$\alpha \leq \frac{\|Ax\|_2^2}{\|x\|_2^2} \leq \beta \tag{4.4}$$

holds for all $x \in \Sigma_k$.

According to (4.4), if we prove that $\dfrac{\|E_{ij}(c)Ax\|_2^2}{\|x\|_2^2}$ possesses an upper bound and a lower bound for all $x \in \Sigma_k$, we can arrive at a conclusion that $E_{ij}(c)A$ satisfies the asymmetric RIP of order $k$.

First, it follows from (4.2) that

$$\|E_{ij}(c)Ax\|_2^2 \leq y_1^2 + \cdots + y_i^2 + \cdots + 2\big((cy_i)^2 + (y_j)^2\big) + \cdots + y_m^2$$

$$\leq 2\big(y_1^2 + \cdots + y_i^2 + \cdots + (cy_i)^2 + (y_j)^2 + \cdots + y_m^2\big)$$

$$= 2\big(y_1^2 + \cdots + (1+c^2)y_i^2 + \cdots + (y_j)^2 + \cdots + y_m^2\big)$$

$$\leq 2(1+c^2)\big(y_1^2 + \cdots + y_i^2 + \cdots + y_j^2 + \cdots + y_m^2\big) = 2(1+c^2)\|Ax\|_2^2.$$

Using (4.4), we can see that $\dfrac{\|E_{ij}(c)Ax\|_2^2}{\|x\|_2^2} \leq 2(1+c^2)\beta$.

Then, we need to prove that

$$\frac{\|E_{ij}(c)A\boldsymbol{x}\|_2^2}{\|\boldsymbol{x}\|_2^2} = \frac{\|A\boldsymbol{x}\|_2^2 - y_j^2}{\|\boldsymbol{x}\|_2^2} + \frac{(cy_i + y_j)^2}{\|\boldsymbol{x}\|_2^2} \tag{4.5}$$

is not infinitely small for all $\boldsymbol{x} \in \Sigma_k$, where the equation follows from (4.3). This can be done by contradiction. So let us assume that there exists some $\boldsymbol{x}_0 \in \Sigma_k$ such that $\dfrac{\|E_{ij}(c)A\boldsymbol{x}\|_2^2}{\|\boldsymbol{x}\|_2^2} \to 0$ as $\boldsymbol{x} \to \boldsymbol{x}_0$. Together with (4.5), this assumption yields

$$\frac{\|A\boldsymbol{x}\|_2^2 - y_j^2}{\|\boldsymbol{x}\|_2^2} = \frac{y_1^2 + \cdots + y_i^2 + \cdots + y_{j-1}^2 + y_{j+1}^2 + \cdots + y_m^2}{\|\boldsymbol{x}\|_2^2} \to 0 \tag{4.6}$$

and

$$\frac{(cy_i + y_j)^2}{\|\boldsymbol{x}\|_2^2} = \left(\frac{cy_i}{\|\boldsymbol{x}\|_2} + \frac{y_j}{\|\boldsymbol{x}\|_2}\right)^2 \to 0. \tag{4.7}$$

Knowing that $\dfrac{y_i^2}{\|\boldsymbol{x}\|_2^2} \to 0$ as a result of (4.6), and that $\dfrac{cy_i}{\|\boldsymbol{x}\|_2} = \pm\sqrt{\dfrac{c^2 y_i^2}{\|\boldsymbol{x}\|_2^2}}$, we can get $\dfrac{cy_i}{\|\boldsymbol{x}\|_2} \to 0$.

Combining it with (4.7), we can see that $\dfrac{y_j}{\|\boldsymbol{x}\|_2} \to 0$ and hence we have $\dfrac{y_j^2}{\|\boldsymbol{x}\|_2^2} \to 0$. Together with (4.6), this yields $\dfrac{\|A\boldsymbol{x}\|_2^2}{\|\boldsymbol{x}\|_2^2} = \dfrac{\|A\boldsymbol{x}\|_2^2 - y_j^2}{\|\boldsymbol{x}\|_2^2} + \dfrac{y_j^2}{\|\boldsymbol{x}\|_2^2} \to 0$, which contradicts (4.4). Therefore, there exists necessarily a constant $\alpha_0$ ($0 < \alpha_0 < \infty$) such that $\dfrac{\|E_{ij}(c)A\boldsymbol{x}\|_2^2}{\|\boldsymbol{x}\|_2^2} \geq \alpha_0$ holds for all $\boldsymbol{x} \in \Sigma_k$. By now we reach the conclusion mentioned above. $\square$

**Theorem 3.6.** *Assume that a matrix A satisfies the RIP of order k. If we transform A into B by executing elementary column operations except column addition, then B still satisfies the RIP of order k.*

*Proof.* We only need to prove that every time when we do an operation of column switching or column multiplication, the RIP order of matrices remains the same. Let $A$ be an $m \times n$ matrix and assume that $A$ satisfies the asymmetric RIP of order $k$.

(i) To swap the positions of two columns of $A$, one can multiply $A$ by the elementary matrix $E_{ij}$ on the right. For any $\boldsymbol{x} \in \Sigma_k$, we have

$$AE_{ij}\boldsymbol{x} = AE_{ij}(x_1, \ldots, x_i, \ldots, x_j, \ldots, x_n)^T = A(x_1, \ldots, x_j, \ldots, x_i, \ldots, x_n)^T.$$

Noting that $E_{ij}\boldsymbol{x} \in \Sigma_k$, and combining it with the assumption that $A$ satisfies (2.4), we get $\alpha\|E_{ij}\boldsymbol{x}\|_2^2 \leq \|AE_{ij}\boldsymbol{x}\|_2^2 \leq \beta\|E_{ij}\boldsymbol{x}\|_2^2$. It is easy to see that $\|E_{ij}\boldsymbol{x}\|_2^2 = \|\boldsymbol{x}\|_2^2$. Thus, we obtain $\alpha\|\boldsymbol{x}\|_2^2 \leq \|AE_{ij}\boldsymbol{x}\|_2^2 \leq \beta\|\boldsymbol{x}\|_2^2$, which shows $AE_{ij}$ satisfies the asymmetric RIP of order $k$.

(ii) To multiply the $i$ th column of $A$ by a nonzero constant $c$, one can multiply $A$ by the elementary matrix $E_i(c)$ ($c \neq 0$) on the right. For any $\boldsymbol{x} \in \Sigma_k$, we have
$$AE_i(c)\boldsymbol{x} = AE_i(c)(x_1, \ldots, x_i, \ldots, x_n)^T = A(x_1, \ldots, cx_i, \ldots, x_n)^T.$$
So we get
$$\|E_i(c)\boldsymbol{x}\|_2^2 = x_1^2 + \cdots + (cx_i)^2 + \cdots + x_n^2. \tag{4.8}$$

Noting that $E_i(c)\boldsymbol{x} \in \Sigma_k$, from the assumption that $A$ satisfies (2.4) we obtain
$$\alpha \|E_i(c)\boldsymbol{x}\|_2^2 \leq \|AE_i(c)\boldsymbol{x}\|_2^2 \leq \beta \|E_i(c)\boldsymbol{x}\|_2^2. \tag{4.9}$$

Due to the value of $c$, we need to consider two situations below.

If $0 < |c| \leq 1$, then it follows from (4.8) that
$$\|E_i(c)\boldsymbol{x}\|_2^2 \leq x_1^2 + \cdots + x_i^2 + \cdots + x_n^2 = \|\boldsymbol{x}\|_2^2,$$
and
$$\|E_i(c)\boldsymbol{x}\|_2^2 \geq c^2 x_1^2 + \cdots + c^2 x_i^2 + \cdots + c^2 x_n^2 = c^2 \|\boldsymbol{x}\|_2^2.$$

Substituting these two inequalities into (4.9) gives us that $c^2 \alpha \|\boldsymbol{x}\|_2^2 \leq \|AE_i(c)\boldsymbol{x}\|_2^2 \leq \beta \|\boldsymbol{x}\|_2^2$.

If $|c| > 1$, then it follows from (4.8) that
$$\|E_i(c)\boldsymbol{x}\|_2^2 > x_1^2 + \cdots + x_i^2 + \cdots + x_n^2 = \|\boldsymbol{x}\|_2^2,$$
and
$$\|E_i(c)\boldsymbol{x}\|_2^2 < c^2 x_1^2 + \cdots + c^2 x_i^2 + \cdots + c^2 x_n^2 = c^2 \|\boldsymbol{x}\|_2^2.$$

Substituting these two inequalities into (4.9) tells us that $\alpha \|\boldsymbol{x}\|_2^2 \leq \|AE_i(c)\boldsymbol{x}\|_2^2 \leq c^2 \beta \|\boldsymbol{x}\|_2^2$.

From the above two situations, we can see that let $\alpha_0 = \min\{\alpha, c^2\alpha\}$ and $\beta_0 = \max\{\beta, c^2\beta\}$, then $\alpha_0 \|\boldsymbol{x}\|_2^2 \leq \|AE_i(c)\boldsymbol{x}\|_2^2 \leq \beta_0 \|\boldsymbol{x}\|_2^2$ holds for all $\boldsymbol{x} \in \Sigma_k$. This shows $AE_i(c)$ satisfies the asymmetric RIP of order $k$. □

## 5. Conclusions

We infer the transformation invariance theorems of the spark, NSP order and RIP order by studying the changes of these three quantitative indexes under elementary transformations of matrices. According to their corollaries, we can get an interesting result that $A\Phi$, as the product of a sensing matrix $A$ satisfying the RIP of order $k$ and an orthonormal basis $\Phi$, does not necessarily satisfy the RIP of order $k$, because $\Phi$ (an invertible matrix) is generally not an invertible diagonal matrix or any permutation of its rows or columns. We already know that if $A$ is a random matrix constructed according to Gaussian or subgaussian distributions, then the product $A\Phi$ still satisfies the RIP of order $k$ with high probability [1, 7, 16]. This property of random matrices is sometimes referred to as *universality* [9, 14]. Obviously, the above result shows that sensing matrices with deterministic constructions can not possess this property.

In the proofs of the transformation invariance theorems, the inferring of inequalities needs us to enlarge or diminish artificially, which makes us can not guarantee that the constants $C$ and $\delta_k$ ($\delta_k$

= $(\beta - \alpha)/(\beta + \alpha)$) in the finally deduced inequalities of the NSP and RIP are the smallest numbers. And we also observed the cases where the final inequalities still hold when we decrease the inferred constants. This means that, after transforming a sensing matrix *A* into *B* by elementary transformations, whether the values of the smallest constants *C* and $\delta_k$ of *B* are the same as *A* is unknown to us. Therefore, whether elementary transformations of matrices have an effect on the smallest constants *C* and $\delta_k$ of sensing matrices is a subject worth exploring in the future.